\documentclass{article}
\usepackage[utf8]{inputenc}

\usepackage{geometry}
\usepackage{wrapfig}
\usepackage{multicol,caption}
\usepackage{titlesec}
\usepackage[T1]{fontenc}
\usepackage{times}
\usepackage{setspace}

\usepackage{amsmath}
\usepackage{latexsym}
\usepackage{amsfonts}
\usepackage{mathrsfs}
\usepackage{upgreek}
\usepackage{graphics}
\usepackage{graphicx}
\usepackage{epstopdf}
\usepackage{epsfig}
\usepackage{amsbsy}
\usepackage{amsfonts}
\usepackage{amsmath}
\usepackage{amssymb}
\usepackage{bm}
\usepackage{color}

\title{An upscaling based three parameter elastic anisotropy model}

\author{E.V. Dontsov}
    
\begin{document}
\maketitle

\abstract{\noindent Rock formations often exhibit transversely anisotropic elastic behavior due to their layered structure. Such materials are characterized by five independent elastic constants. In the context of petroleum applications, it is often challenging to accurately measure all these elastic parameters. At the same time, the effect of elastic anisotropy can be noticeable and therefore there is a need to include it in some form. To fill the gap, this study proposes a three parameter elastic anisotropy model. It captures the dominant anisotropic behavior and yet has only three elastic constants that are relatively easy to measure in laboratory. The approach is based on upscaling of a periodically layered material characterized by equal height layers with different Young's moduli and the same Poisson's ratio. The resultant upscaled material is transversely isotropic and is also physically admissible. The developed approach allows to effectively reconstruct or estimate the values of the two remaining parameters needed for the transversely isotropic model. Comparison between the reconstructed results and the measured values are compared for several rock types. }


\section{Introduction}

Sedimentary and layered nature of rock formations often leads to transversely isotropic elastic behavior. One of the early works includes~\cite{Thom1986}, in which weak anisotropy is investigated and the so-called Thomsen parameters are introduced to describe it. The study~\cite{John1994} reports the values for the anisotropic elastic constants for Devonian-Missisipian shale that are obtained using sound velocity measurements and the variation of the aforementioned elastic parameters versus confining pressure is observed. A comparison between static and dynamic parameters is investigated in~\cite{PenaMS1998}. The dynamic elastic constants are measured by utilizing the variation of sound speed with respect to bedding orientation, while the static properties are calculated on the basis of the stress-strain response of core plugs with various orientations. There are many more other authors who investigated anisotropic elastic properties of shales, see e.g.~\cite{Perv2008,Kuila2011,Nadri2011,Sond2011,Josh2012,Jin2018,Yan2019,Yuri2020}. Some use dynamic velocity measurements and some focus on the static experiments. It is also interesting to mention the study~\cite{Josh2012}, in which correlations between petrophysical and geomechanical properties are obtained. At the same time, authors in~\cite{Yan2019} developed correlations between anisotropic elastic constants based on the obtained data. The study~\cite{Jin2018} also reported the anisotropic values of fracture toughness for Marcellus shale, but surprisingly the degree of anisotropy is relatively mild, on the order of 10\%. Finally, the study~\cite{Yuri2020} investigated compaction trends for shales and systematically analyzed the anisotropy of various clays that appear in shales.

Elastic anisotropy of rock formations is important for hydraulic fracture propagation. The vast majority of hydraulic fracture simulators assume that the rock formation is isotropic, even though it can be layered. One of the first articles devoted to the influence of elastic anisotropy on hydraulic fracture propagation is~\cite{Chert2012}, where the effect of anisotropy is investigated in the context a constant height model. It is concluded that for such a geometry the solution is identical to that for an isotropic material, but features an apparent elastic modulus that depends on the anisotropic constants. A plain strain fracture is analyzed in~\cite{Laub2014a}. It is again observed that the solution is the same as for an isotropic material, but the apparent elastic modulus depends on the anisotropic constants as well as on orientation of the fracture relative to the bedding layers. The analysis is extended to a uniformly pressurized elliptical fracture in~\cite{Laub2014b}. The variation of the elliptical fracture aspect ratio versus propagation regime and the degree of anisotropy is investigated in~\cite{BessmARMA2018}. Further, the study~\cite{Dont2018e} outlines the parametric map for an elliptical hydraulic fracture propagating in a homogeneous anisotropic material. Numerical modeling is employed in~\cite{Sesetty2018} and~\cite{Sesetty2018b} to better understand the influence of elastic anisotropy on the hydraulic fracture propagation near the wellbore and for multiple pseudo-3D fractures. Finally, results of the modeling with a fully planar hydraulic fracture model are presented in~\cite{MoukARMA1019}. 

In order to specify the transversely isotropic material for the purpose of hydraulic fracture modeling, five elastic constants should be provided. On the other hand, isotropic material is quantified by only two constants. Therefore, there can be two more intermediate models with three and four constants. An example of the four parameter model can be found in~\cite{Elita2019}, while the first three parameter model is introduced in~\cite{Schoen1996} and is called ANNIE. There are two modifications to ANNIE, presented in~\cite{Quirein2014} and~\cite{Murphy2015}, respectively. Also, another three parameter model is effectively used in~\cite{BessmARMA2018}, where the result is obtained by upscaling a periodically layered material in which the layers have different Young's modulus and the same Poisson's ratio. The primary use of the simplified three parameter models is to tackle often ocurring data poor cases for which values of all five elastic constants are not available. As a result, the purpose of this study is to evaluate and to compare different three parameter elastic models in the context of hydraulic fracture modeling.

\section{Relations between elastic constants for a transversely isotropic elastic material}

As a starting point, transversely isotropic elastic material needs to be defined. This is a material, which is isotropic in the horizontal plane, but the behavior in the vertical direction is different. Hooke's law for such a material can be written as
\begin{equation}\label{Hookeslaw}
\begin{bmatrix} \sigma_{xx}\\ \sigma_{yy}\\ \sigma_{zz}\\ \sigma_{xz} \\ \sigma_{yz} \\ \sigma_{xy}\end{bmatrix} =\begin{bmatrix} C_{11}&C_{12}&C_{13}&0&0&0\\ C_{12}&C_{11}&C_{13}&0&0&0\\ C_{13}&C_{13}&C_{33}&0&0&0\\ 0&0&0&C_{44}&0&0 \\ 0&0&0&0&C_{44}&0 \\ 0&0&0&0&0&(C_{11}\!-\!C_{12})/2\end{bmatrix} \begin{bmatrix} \varepsilon_{xx}\\ \varepsilon_{yy}\\ \varepsilon_{zz}\\ 2\varepsilon_{xz} \\ 2\varepsilon_{yz} \\ 2\varepsilon_{xy}\end{bmatrix},
\end{equation}
where $\sigma_{ij}$ are the components of the stress tensor, $\varepsilon_{ij}$ are the components of the strain tensor, and $C_{ij}$ are the elastic or stiffness constants. Here it is assumed that the vertical axis is $z$, while the horizontal $(x,y)$ plane is is the plane of isotropy. Note that there are five independent elastic constants, namely $C_{11}$, $C_{12}$, $C_{13}$, $C_{33}$, and $C_{44}$. It is also useful to write the Hooke's law~(\ref{Hookeslaw}) in terms of compliances as 
\begin{equation}\label{Hookeslaw2}
\begin{bmatrix} \varepsilon_{xx}\\ \varepsilon_{yy}\\ \varepsilon_{zz}\\ 2\varepsilon_{xz} \\ 2\varepsilon_{yz} \\ 2\varepsilon_{xy}\end{bmatrix}=\begin{bmatrix} S_{11}&S_{12}&S_{13}&0&0&0\\ S_{12}&S_{11}&S_{13}&0&0&0\\ S_{13}&S_{13}&S_{33}&0&0&0\\ 0&0&0&S_{44}&0&0 \\ 0&0&0&0&S_{44}&0 \\ 0&0&0&0&0&2(S_{11}\!-\!S_{12})\end{bmatrix} \begin{bmatrix} \sigma_{xx}\\ \sigma_{yy}\\ \sigma_{zz}\\ \sigma_{xz} \\ \sigma_{yz} \\ \sigma_{xy}\end{bmatrix},
\end{equation}
where $S_{ij}$ are the compliance coefficients. The relationships between the compliance and stiffness constants follow the matrix inversion rules and can be summarized as
\begin{eqnarray}\label{Sij}
&&S_{11} = \dfrac{1}{\Delta}\dfrac{C_{11}\!-\!C_{13}^2/C_{33}}{C_{11}\!-\!C_{12}},\qquad S_{12} = -\dfrac{1}{\Delta}\dfrac{C_{12}\!-\!C_{13}^2/C_{33}}{C_{11}\!-\!C_{12}},\qquad S_{13} = -\dfrac{1}{\Delta}\dfrac{C_{13}}{C_{33}},\notag\\
&& S_{33} = \dfrac{1}{\Delta}\dfrac{C_{11}\!+\!C_{12}}{C_{33}},\qquad S_{44} = \dfrac{1}{C_{44}},\qquad \Delta = C_{11}\!+\!C_{12}\!-\!2C_{13}^2/C_{33}.
\end{eqnarray}
For the future reference, it is also useful to provide alternative definitions of the elastic constants that is commonly used in engineering applications
\begin{equation}\label{EhEv}
S_{11}=\dfrac{1}{E_h},\qquad S_{12} = -\dfrac{\nu_h}{E_h},\qquad S_{13} = -\dfrac{\nu_v}{E_v},\qquad S_{33} = \dfrac{1}{E_v},\qquad S_{44} = \dfrac{1}{G_{vh}}.
\end{equation}
Here $E_h$ is the horizontal Young's modulus, $\nu_h$ is the horizontal Poisson's ratio, $E_v$ is the vertical Young's modulus, $\nu_v$ is the vertical Poisson's ratio, and $G_{vh}$ is the shear modulus defined in either $(x,z)$ or $(y,z)$ plane.

\section{Three parameter elastic model}

\begin{figure}
\begin{centering}
\includegraphics[width=0.5\columnwidth]{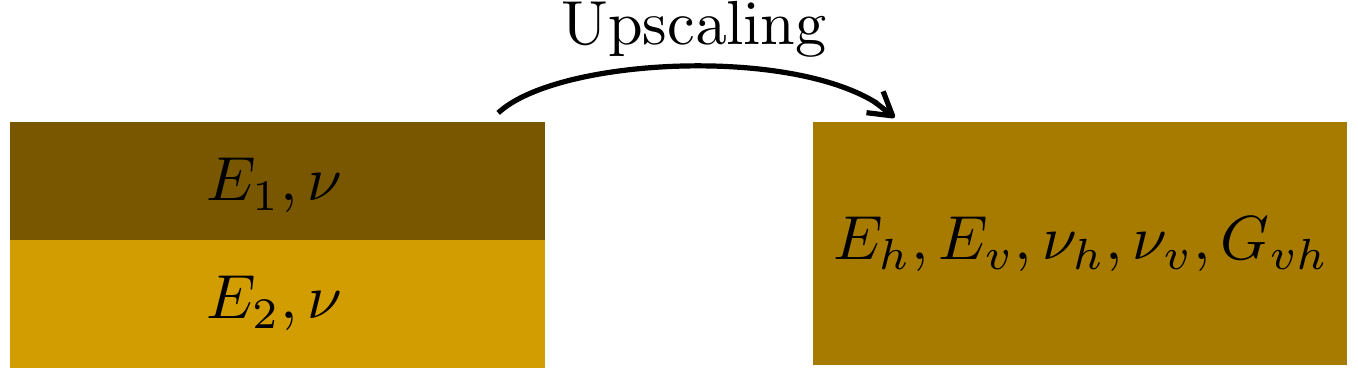}
\par\end{centering}
\caption{Upscaling of a periodically layered material.\label{fig1}}
\end{figure}

To develop the three parameter anisotropic model, the concept of upscaling is applied to a layered material. In particular, as shown in Fig.~\ref{fig1} the problem consists of a periodically layered material, in which the layers have equal height and are characterized by $E_1$, $\nu$ and $E_2$, $\nu$. In other words, the layers have different Young's moduli and the same Poisson's ratio. Thus, there are only three independent parameters in the model, namely, $E_1$, $E_2$, and $\nu$. Note that such a model was considered in~\cite{BessmARMA2018} for the purpose of analyzing aspect ratio of a hydraulic fracture. By applying Backus averaging or upscaling~\cite{Back1962}, the elastic constants $C_{ij}$ can be calculated as
\begin{eqnarray}\label{CijBackus}
&&C_{11} = \dfrac{1}{1\!-\!\nu^2}\langle E\rangle + \dfrac{\nu^2}{(1\!-\!\nu^2)(1\!-\!2\nu)}\langle E^{-1}\rangle^{-1},\qquad C_{12} = \dfrac{\nu}{1\!-\!\nu^2}\langle E\rangle + \dfrac{\nu^2}{(1\!-\!\nu^2)(1\!-\!2\nu)}\langle E^{-1}\rangle^{-1},\notag\\
&& C_{13} = \dfrac{\nu}{(1\!+\!\nu)(1\!-\!2\nu)}\langle E^{-1}\rangle^{-1},\qquad  C_{33} = \dfrac{1\!-\!\nu}{(1\!+\!\nu)(1\!-\!2\nu)}\langle E^{-1}\rangle^{-1},\qquad C_{44} = \dfrac{1}{2(1\!+\!\nu)}\langle E^{-1}\rangle^{-1}.
\end{eqnarray}
Here $\langle E\rangle = (E_1\!+\!E_2)/2$ is the arithmetic average of the Young's moduli, while $\langle E^{-1}\rangle^{-1} = 2E_1E_2/(E_1\!+\!E_2)$ is the geometric average of the Young's moduli. Thus, equations~(\ref{CijBackus}) provide the values for all five elastic constants, but there are only three independent parameters. The primary advantage of using upscaling to define the remaining two parameters is the fact that the resultant anisotropic material is guaranteed to be physically admissible, i.e. the stiffness matrix in~(\ref{Hookeslaw}) is positive definite. In addition, many of the reservoir rocks are finely layered and individual layers often have very different Young's moduli, while the Poisson's ratio does not change that significantly. This observation provides an additional justification for the model.

The answer presented in~(\ref{CijBackus}) depends on the properties of individual layers, i.e. on $E_1$, $E_2$, and $\nu$. For engineering applications, it is instructive to reformulate the result in terms of parameters that are can be measured in a laboratory. In particular, it is assumed that the values of $E_h$, $\nu_h$, and $E_v$ are known, and the goal is to express the values of $\nu_v$ and $G_{vh}$ in terms of the known parameters. The procedure for this is the following. The upscaled values of $C_{ij}$~(\ref{CijBackus}) need to be substituted into~(\ref{Sij}) to find $S_{ij}$ in terms of $E_1$, $E_2$, and $\nu$. After that, the relations~(\ref{EhEv}) are used to relate the $S_{ij}$ to the engineering parameters. The result can be summarized as
\begin{eqnarray}\label{Sijupscaling}
&&\Delta = \dfrac{1}{1\!-\!\nu}\langle E\rangle,\qquad 
S_{11}=\dfrac{1}{E_h}=\dfrac{1}{\langle E\rangle},\qquad 
S_{12}=S_{13}=-\dfrac{\nu_h}{E_h}=-\dfrac{\nu_v}{E_v}=-\dfrac{\nu}{\langle E\rangle},\notag\\
&&S_{33}=\dfrac{1}{E_v}=\dfrac{(1\!+\!\nu)(1\!-\!2\nu)}{(1\!-\!\nu)\langle E^{-1}\rangle^{-1}} + \dfrac{2\nu^2}{(1\!-\!\nu)\langle E\rangle},\qquad
S_{44}=\dfrac{1}{G_{vh}}= \dfrac{2(1\!+\!\nu)}{\langle E^{-1}\rangle^{-1}}.
\end{eqnarray}
From equations~(\ref{Sijupscaling}) it becomes apparent that $\langle E\rangle=E_h$ and $\nu=\nu_h$, while the vertical Poisson's ratio becomes
\begin{equation}\label{nuv}
\nu_v = \nu_h\dfrac{E_v}{E_h}.
\end{equation}
The harmonic average of the moduli can be expressed as
\begin{equation}\label{Eharmav}
\langle E^{-1}\rangle^{-1} = \dfrac{(1\!+\!\nu_h)(1\!-\!2\nu_h)}{1\!-\!\nu_h}\biggl[\dfrac{1}{E_v}- \dfrac{2\nu_h^2}{(1\!-\!\nu_h)E_h}\biggr]^{-1}.
\end{equation}
And finally the expression for the shear modulus is
\begin{equation}\label{Gvh}
G_{vh}= \dfrac{(1\!-\!2\nu_h)}{2(1\!-\!\nu_h)}\biggl[\dfrac{1}{E_v}- \dfrac{2\nu_h^2}{(1\!-\!\nu_h)E_h}\biggr]^{-1}=\dfrac{E_h}{2(1\!+\!\nu_h)}\biggl[ 1+\dfrac{1\!-\!\nu_h}{(1\!+\!\nu_h)(1\!-\!2\nu_h)}\Bigl(\dfrac{E_h}{E_v}-1\Bigr) \biggr]^{-1}.
\end{equation}
To summarize, given the values of $E_h$, $E_v$, and $\nu_h$, the three parameter anisotropy model provides the values for $\nu_v$~(\ref{nuv}) and $G_{vh}$~(\ref{Gvh}). Note that $E_v\leqslant E_h$ and $\nu_h<0.5$, therefore the value of the shear modulus is always positive.

It is instructive to compare the obtained result to the two existing models called ANNIE~\cite{Schoen1996} and MANNIE~\cite{Quirein2014}. In the former model, it is assumed that $C_{12}=C_{13}=C_{33}\!-\!2C_{44}$, see~\cite{Schoen1996}. As a result, the relationships between the $C_{ij}$ and engineering constants can be summarized as
\begin{equation}\label{ANNIE1}
C_{11} = \dfrac{1}{\Delta_S}\dfrac{1-\nu_v^2E_h/E_v}{1+\nu_h},\qquad C_{11}-C_{12} = \dfrac{E_h}{1+\nu_h},\qquad C_{33} = \dfrac{1}{\Delta_S}\dfrac{1-\nu_h}{E_h}E_v,\qquad C_{12} = \dfrac{\nu_v}{\Delta_S},\qquad 
\end{equation}
where $\Delta_S = (1\!-\!\nu_h)/E_h-2\nu_v^2/E_v$. The above equations can be combined to obtain
\begin{equation}\label{nuvGvhANNIE}
\nu_v = \dfrac{(1+\nu_h)-\sqrt{(1+\nu_h)^2-4\nu_hE_h/E_v}}{2E_h/E_v},\qquad G_{vh} =\dfrac{1}{2\Delta_S}\biggl[(1\!-\!\nu_h)\dfrac{E_v}{E_h}-\nu_v\biggr].
\end{equation}
MANNIE model assumes $C_{13}=K_2C_{12}$ and $C_{11}=K_1(C_{11}\!-\!C_{12}\!-\!2C_{44}\!+\!C_{33})$~\cite{Quirein2014}, where $K_1=1.1$ and $K_2=0.8$. This leads to the very similar relations between the elastic constants
\begin{equation}\label{MANNIE1}
C_{11} = \dfrac{1}{\Delta_S}\dfrac{1-\nu_v^2E_h/E_v}{1+\nu_h},\qquad C_{11}-C_{12} = \dfrac{E_h}{1+\nu_h},\qquad C_{33} = \dfrac{1}{\Delta_S}\dfrac{1-\nu_h}{E_h}E_v,\qquad K_2C_{12} = \dfrac{\nu_v}{\Delta_S},
\end{equation}
where $\Delta_S = (1\!-\!\nu_h)/E_h-2\nu_v^2/E_v$. 
After some algebraic manipulations, the resultant expressions for $\nu_v$ an $G_{vh}$ are
\begin{equation}\label{nuvGvhMANNIE}
\nu_v = \dfrac{(1+\nu_h)/K_2-\sqrt{(1+\nu_h)^2/K_2^2-4\nu_hE_h/E_v}}{2E_h/E_v},\qquad G_{vh} = \dfrac{1}{2}\bigl(C_{11}(1\!-\!K_1^{-1})\!-\!C_{12}\!+\!C_{33}\bigr),
\end{equation}
where the expressions for $C_{ij}$ from~(\ref{MANNIE1}) should be used to calculate $G_{vh}$ from the above equation.

The results in~(\ref{nuvGvhANNIE}) and~(\ref{nuvGvhMANNIE}) demonstrate that such formulations can potentially lead to complex valued elastic constants, which is unphysical. In particular the maximum allowable ratio between the Young's moduli is
\begin{equation}\label{maxErat}
\max\Bigl\{\dfrac{E_h}{E_v}\Bigr\} = \dfrac{(1+\nu_h)^2}{4\nu_hK_2^2}.
\end{equation}
The result with $K_2=0.8$ applies to MANNIE model, while $K_2=1$ for ANNIE model. For a typical value of $\nu_h=0.2$, the maximum allowable level of Young's moduli anisotropy is approximately 1.8 for ANNIE model and 2.8 for MANNIE model. Such a restriction significantly limits applicability of these three parameter models.

\begin{figure}
\begin{centering}
\includegraphics[width=0.65\columnwidth]{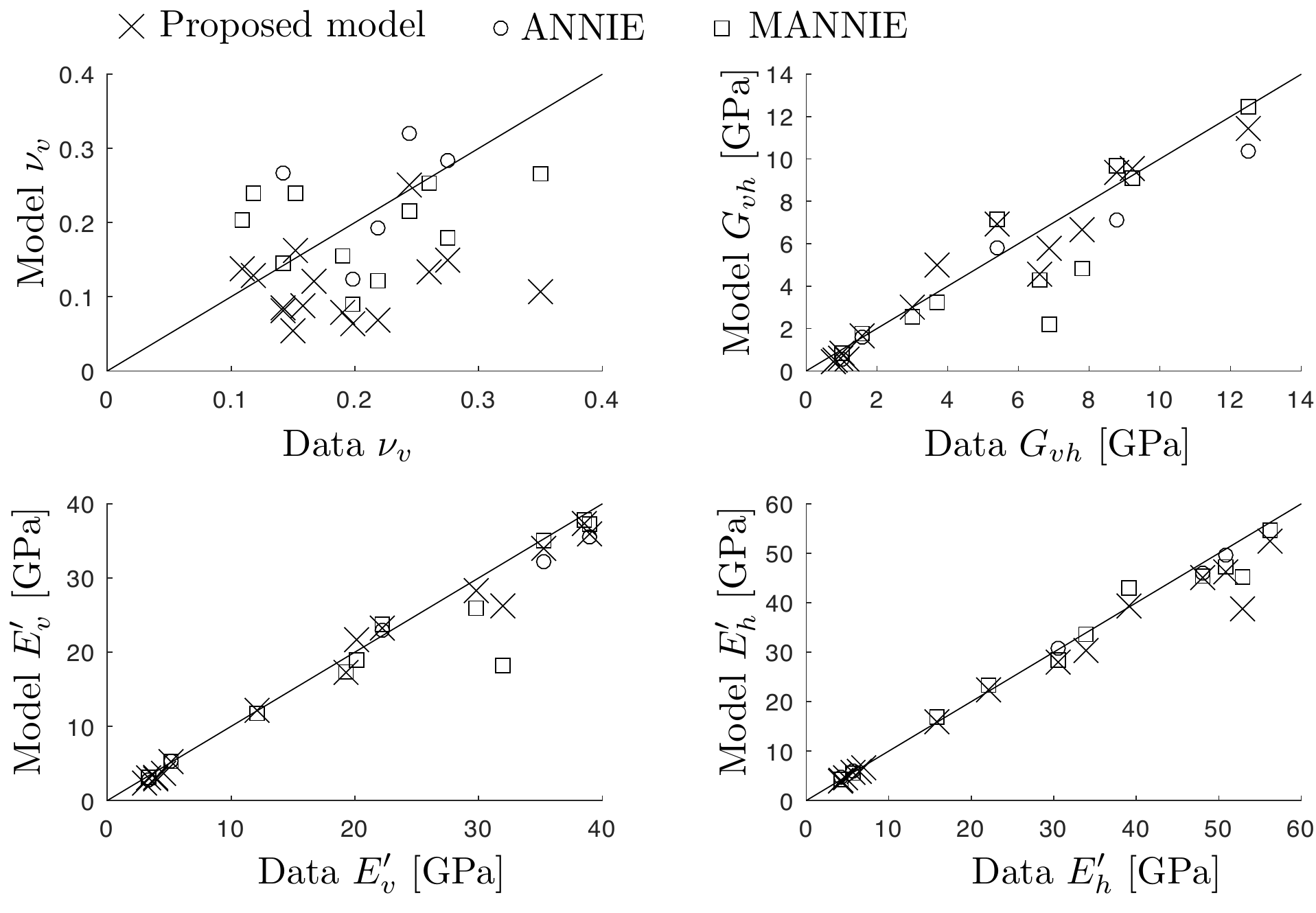}
\par\end{centering}
\caption{Comparison between the predictions of various three parameter anisotropic models and actual data.\label{fig2}}
\end{figure}

To demonstrate predictability of the proposed model, several data points for shales are taken from~\cite{Chert2012,Laub2014a,Jin2018,John1994,Sond2011}. All the data points have five elastic constants. It is assumed that $E_h$, $E_v$, and $\nu_h$ are given, while the values for $\nu_v$ and $G_{hv}$ are predicted by the three parameter models. Fig.~\ref{fig2} plots a comparison between the predicted values of $\nu_v$ and $G_{hv}$ against their respective true values. Three models are compared: the proposed model~(\ref{nuv}) and~(\ref{Gvh}), ANNIE model~(\ref{nuvGvhANNIE}), and MANNIE model~(\ref{nuvGvhMANNIE}). For the latter two models the data points corresponding to complex valued predictions are ignored. In addition, to cater for hydraulic fracture applications, the apparent moduli $E_v'$ and $E_h'$ are compared. These are defined as
\begin{equation}\label{EhpEvp}
E'_h = \dfrac{C_{11}^2-C_{12}^2}{C_{11}},\qquad E'_v=2\biggl(\dfrac{C_{33}}{C_{11}C_{33}-C_{13}^2}\Bigl(\dfrac{1}{C_{44}}+\dfrac{2}{C_{13}\!+\!\sqrt{C_{11}C_{33}}}\Bigr)\biggr)^{-1/2}.
\end{equation}
The modulus $E_h'$ is the apparent elastic modulus for a horizontally oriented plane strain fracture, while $E_v'$ is the corresponding apparent modulus for a vertically oriented plane strain fracture, see e.g.~\cite{Laub2014b,BessmARMA2018,Dont2019}. The values of these parameters determine hydraulic fracture growth and therefore are the relevant moduli that need to be predicted accurately.

Results shown in Fig.~\ref{fig2} demonstrate that the prediction of $\nu_v$ is the least accurate and the accuracy of prediction is approximately the same for all the models considered. At the same time, the predictions of $G_{vh}$, $E'_v$, and $E'_h$ are more accurate, which again applies for all models. The main conclusion is therefore the following. All three models are able to reasonably approximate the remaining elastic constants for the considered data. At the same time, given the specificity of prescribing $E_h$, $E_v$, and $\nu_h$, both ANNIE and MANNIE models have a very limited applicability range in terms of the anisotropy ratio $E_h/E_v$. This makes the proposed upscaling based model more suitable as a general three parameter anisotropic model.

\section{Summary}

This study proposes a three parameter elastic anisotropy model. The approach is based on upscaling of a periodically layered material with different Young's moduli and the same Poisson's ratio. The input parameters for the model are the horizontal Young's modulus and Poisson's ratio, as well as the vertical Young's modulus. Mathematical expressions for calculating the vertical Poisson's ratio and the shear modulus in the vertical plane are presented. To validate the model, data for a series of measurements for shale samples is gathered from the literature. The predicted values of the vertical Poisson's ratio and the shear modulus are compared to the actual values and show a good level of agreement. In addition, the developed model is compared to the existing three parameter anisotropic models ANNIE and MANNIE. Both of the latter models feature similar level of accuracy compared to the proposed model. However, these models predict unphysical results when the ratio between the horizontal and vertical Young's modulus exceeds a certain value. Therefore, such models are less suitable for the role of a general three parameter elastically anisotropic model.



\end{document}